\documentclass{gAPA2e}
\usepackage{times,amsfonts,amsmath,amssymb,amsbsy,graphicx}
\usepackage{mathbbol,epsfig,subfigure,color} 
\advance\parskip 4pt
\usepackage{epstopdf}

\def\half{{\textstyle\frac12}}
\def\sech{\mathop{\rm sech}\nolimits}
\let\@=\mathbf
\def\sf#1#2{{\textstyle\frac{#1}{#2}}}
\def\Real{{\mathbb{R}}}

\begin{document}
\title{\bf On the construction of the KP line-solitons and their interactions}
\author{Sarbarish Chakravarty$^{a,1}$,\, Tim Lewkow$^a$ and Ken-ichi Maruno$^b$ \\[1ex]
\small\it\ 
$^a$Department of Mathematics, University of Colorado, Colorado Springs, CO 80933\\
\small\it\ 
$^b$Department of Mathematics, University of Texas-Pan American, Edinburg, TX 78539 }

\maketitle
\begin{abstract}
\noindent The line-soliton solutions of the  Kadomtsev--Petviashvili (KP) equation are
investigated in this article using the $\tau$-function formalism. In particular, 
the Wronskian and the Grammian forms of the $\tau$-function are discussed, and the
equivalence of these two forms are established. Furthermore, the interaction properties
of two special types of $2$-soliton solutions of the KP equation are studied in details.

\medskip


\medskip

\end{abstract}

\footnotetext[1]{Corresponding author. Email:\, chuck@math.uccs.edu}


\section{Introduction}
\noindent The eponymous Kadomtsev-Petviashvili (KP) equation 
\begin{equation}
(-4u_t+ u_{xxx} +6uu_x)_x + 3\beta u_{yy} = 0 \,,
\label{kp}
\end{equation}
discovered in 1970, describes the dynamics of small-amplitude, 
long wavelength, solitary waves in two 
dimensions ($xy$-plane)~\cite{KP:70}, and arises in the study of
water waves, plasma and various other areas of physical significance
(see e.g.,~\cite{IR:00} for a review).
In equation \eqref{kp}, $u=u(x,y,t)$ is the wave amplitude, the subscripts 
denote partial derivatives with respect to $x, y, t$, and $\beta = \pm 1$.
Throughout this article, equation \eqref{kp} with $\beta=1$, 
which corresponds to the negative-dispersion KP equation (KP II) will be 
considered, and will be referred to as the KP equation.

The KP equation is a completely integrable system whose underlying mathematical
properties have been extensively studied during the last four decades. 
They are well documented in several monographs including 
(but not limited to)~\cite{NMPZ:84, AC:91, D:91, H:04}. 
These properties include the existence of multi-soliton and periodic solutions and
the Lax representation of the inverse scattering transform. A major breakthrough in the 
KP theory occurred in 1981 when Sato \cite{S:81} formulated the KP equation in terms 
of an infinite dimensional Grassmann manifold 
known as the Sato universal Grassmannian.
A finite dimensional version of the Sato theory corresponding to the real 
Grassmannian Gr$(N,M)$ (the set of $N$-dimensional subspaces of $\mathbb{R}^M$)
leads to a simple algebraic construction of a special class of solitary wave 
solutions of the KP equation called the {\em line-soliton} solutions. These are 
real, non-singular solutions 
which decay exponentially in the $xy$-plane except along certain directions.
Specifically, such a solution is localized along two distinct sets of rays 
(referred to as line solitons) as $y \to \pm \infty$, and form spatial interaction 
patterns in the finite region of the $xy$-plane. 

The simplest example of a 
line-soliton is the one-soliton solution of KP given by
\begin{equation}
u(x,y,t)= \alpha\sech^2(\@k\cdot\@r+\omega t + \theta_0) \,,
\label{onesoliton}
\end{equation}
which is a traveling wave with $\@r:=(x,y)$, amplitude $\alpha=\half(k_2-k_1)^2$,
wave vector $\@k:=(k_x,\,k_y) = \half(k_2-k_1,\,k_2^2-k_1^2)$ and
frequency $\omega = \half(k_2^3-k_1^3)$, where $k_1,\,k_2$ are distinct real parameters
with $k_1 < k_2$. Clearly, the soliton amplitude depends on the wave vector $\@k$, 
which together with the frequency $\omega$ satisfy the soliton 
dispersion relation
$$ 4\omega k_x=4k_x^4+3k_y^2 \,. $$
The solitary wave-form given by \eqref{onesoliton} is localized in the $xy$-plane 
along a line which makes an angle $\psi$ measured counterclockwise from the $y$-axis 
where 
$$
\tan \psi = k_y/k_x = k_1+k_2 \,, \qquad \qquad 
-\sf{\pi}{2} < \psi < \sf{\pi}{2}\,.$$
A one-soliton solution is shown in Figure 1(a).
Since the one-soliton solution is characterized by two real parameters 
$k_1, \, k_2$, it is convenient to denote this solution simply as the $[1,2]$-soliton.
Note that when $k_1+k_2=0$, the solution
in \eqref{onesoliton} becomes $y$-independent and reduces
to the one-soliton solution of the Korteweg-de~Vries (KdV) equation.

The soliton interactions of the KP equation were originally described via 
a 2-soliton solution with a ``X"-shape pattern in the plane formed by the
intersection of two line solitons but with a parallel shift of the two 
lines at the intersection (the phase shift). This 2-soliton solution 
(see Figure 1(c)) referred to as the O-type soliton 
(``O" stands for {\it original}) and its $N$-soliton generalization
were obtained independently by several authors using integral 
equations~\cite{ZS:74} and other algebraic methods~\cite{S:76,FN:83}.
In 1977, Miles \cite{M:77} pointed out that the O-type 
2-soliton solution becomes singular if the angle of the intersection is smaller
than a certain critical value. As the angle approaches the critical value from above, 
the 2-soliton phase shift tends to infinity, and at the critical angle the 
O-type solution degenerates to a ``Y-shape" with only three line solitons
interacting resonantly (see also \cite{NR:77}).
It turns out that such Y-shape interacting wave-forms are also exact solutions of 
the KP equation~\cite{F:80,OW:83}.  Apart from the ones mentioned thus far, 
no other soliton solutions of the KP equation were known for quite some time until
recently when more general types of resonant and non-resonant line-soliton solutions
were reported in several works including ~\cite{BPPP:01, M:02, BK:03, PF:05}.
During the last 5 years, considerable progress has been made toward the problem of 
classifying all exact line-soliton solutions of the KP 
equation~\cite{K:04, BC:06, CK:08, CK:08b}. These studies have revealed a large variety 
of soliton solutions which were totally overlooked in the past.
Generically, these solutions of KP consist of two distinct sets of line solitons 
of different amplitudes and along different directions in the $xy$-plane as 
$y \to \pm \infty$. Locally, each line soliton denoted as the $[i,j]$-soliton has the 
form of a one-soliton solution as in \eqref{onesoliton}, parametrized by two 
distinct real parameters $k_i < k_j, \, i< j$. The soliton amplitude 
is given by $\alpha_{ij}=\half(k_j-k_i)^2$ and the soliton angle $\psi_{ij}$ 
(counterclockwise from the $y$-axis) satisfies $\tan \psi_{ij} = k_i+k_j$. 

\begin{figure}[t!]
\begin{center}
\subfigure[]{
\resizebox*{4cm}{!}{\includegraphics{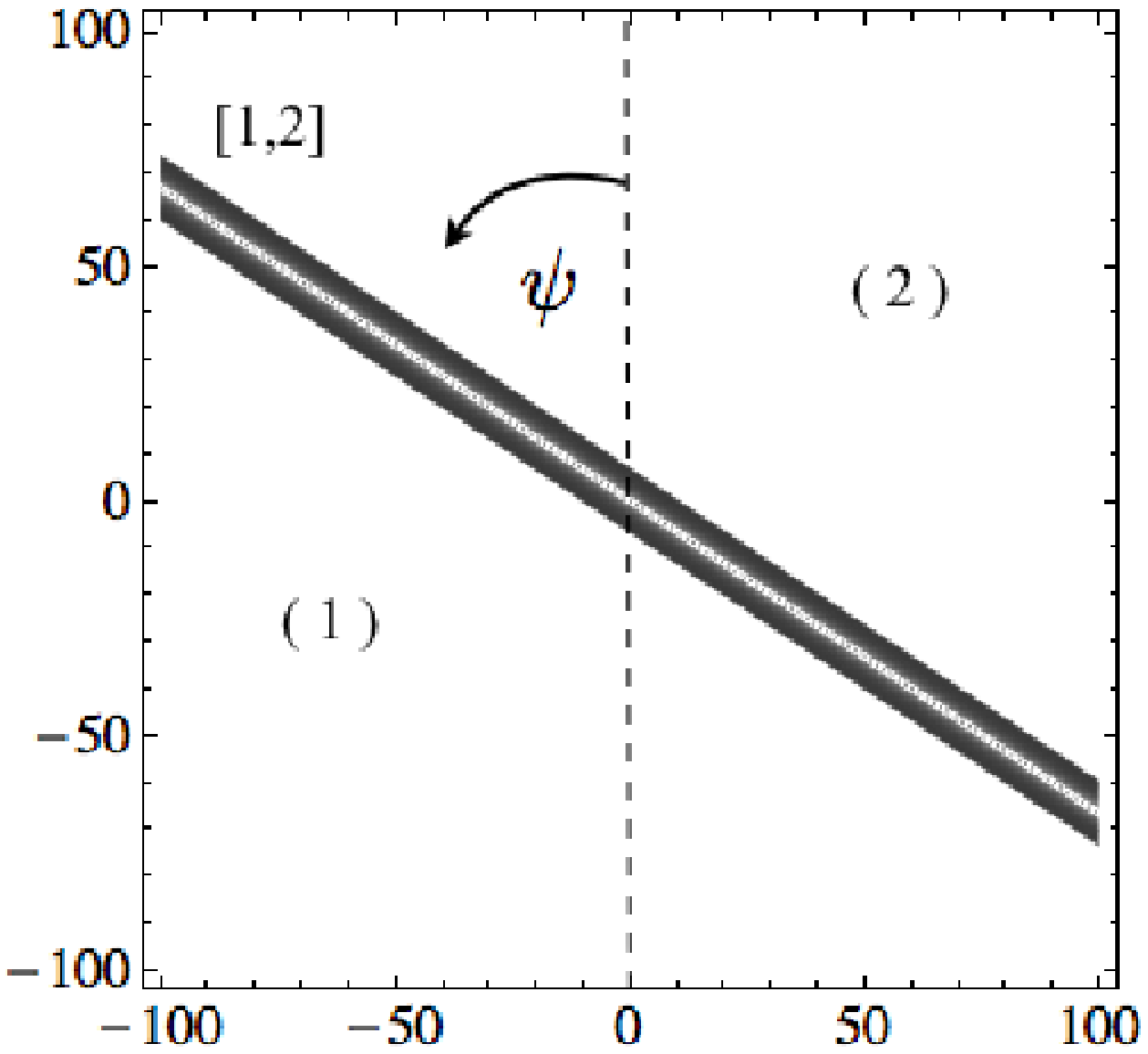}}} \hskip 0.35 cm 
\subfigure[]{
\resizebox*{4cm}{!}{\includegraphics{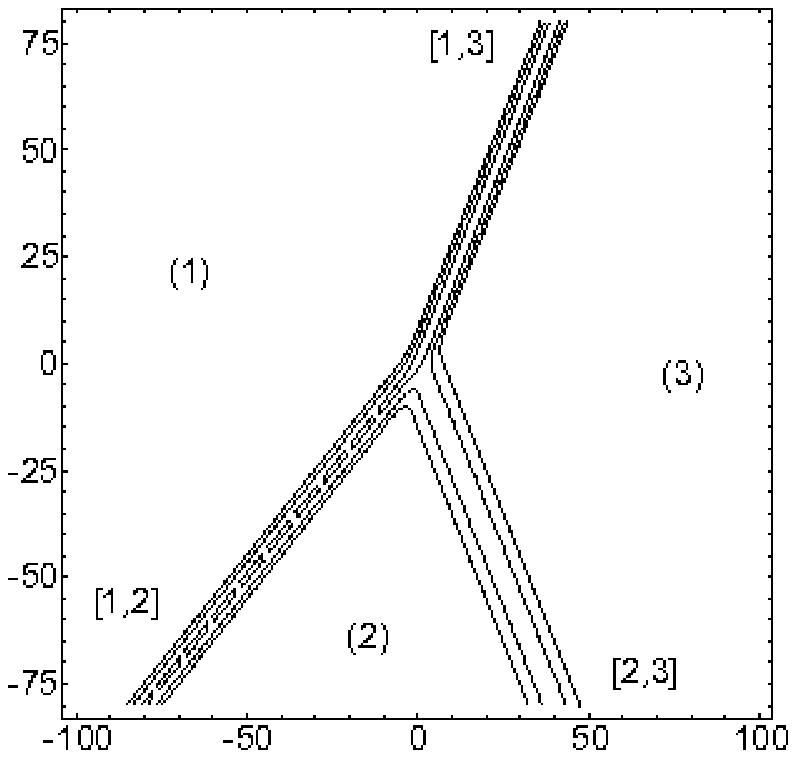}}} \hskip 0.35 cm 
\subfigure[]{
\resizebox*{4cm}{!}{\includegraphics{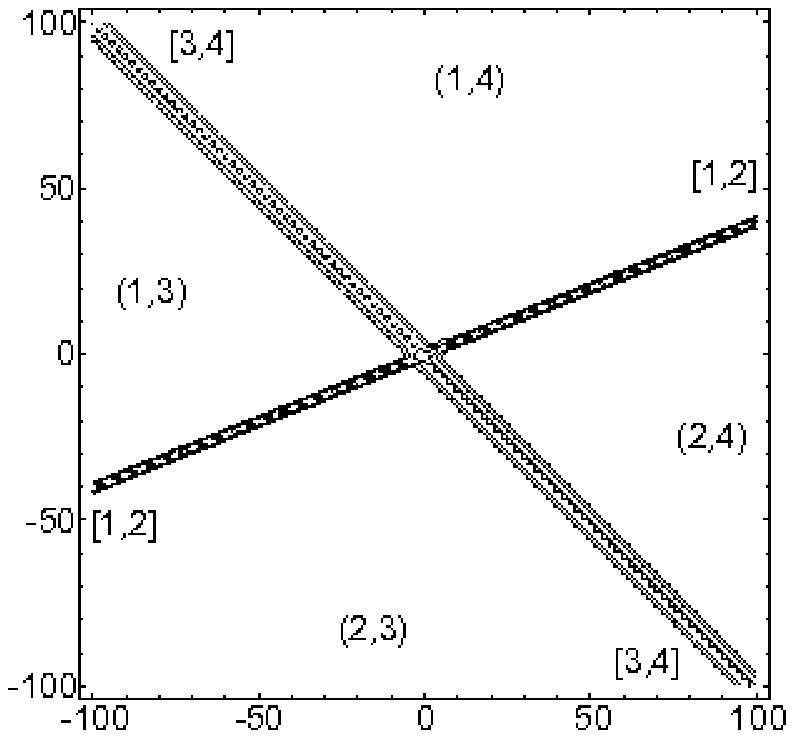}}} 
\caption{Line-soliton solutions of KP:\,(a)~a one-soliton solution with 
$(k_1,k_2)=(0.5, 1)$ at $t=0$, $\psi$ is the soliton angle; 
(b)~ a Miles's Y-shape soliton solution with $(k_1,k_2,k_3)=(-1,0,\frac12)$ at $t=0$;
(c)~an O-type 2-soliton with $(k_1,\dots,k_4)=(-1,-\frac12,\frac12,2)$
 at $t=0$. In each figure the dominant exponentials in the $\tau$-function (see section 2)
 in different regions
 of the $xy$-plane are labeled by the indices in parentheses,
 and the asymptotic line solitons are labeled by the index pair in square brackets.}
\label{fig1}
\end{center}
\end{figure}

There are several direct and algebraic methods to construct the KP line-soliton solutions 
which are usually derived from either a Wronskian or 
from a Gram determinant (Grammian). In principle any given 
soliton solution can be represented in either form,
although this fact has been explicitly shown for only the one-soliton and the O-type
soliton solutions. For more general line-soliton solutions, it is often more difficult
to impose regularity conditions on the solution when it is represented in one particular
form than the other, thus rendering one solution generating method less efficient
than another. In this note, we discuss both the Wronskian and the Grammian forms of the
general line-soliton solutions, and establish the equivalence between the two 
representations in an explicit fashion. That is, we show how to derive one form of
the solution from the other, and vice-versa.
Another purpose of this paper is to describe the interaction properties of the various
types of line-soliton solutions. In particular, we present a detailed discussion of two
distinct types of two-soliton solutions both of which form an ``X"-shape pattern 
on the $xy$-plane but interact in a significantly different way. 

The paper is planned as follows. In Section 2, we introduce the $\tau$-function
of the KP equation and present its Wronskian and Grammian forms generating the line-soliton
solutions. Then the equivalence of these two representations of the $\tau$-function is 
established. Section 3 is devoted to a brief description of the distinct types of
the 2-soliton solutions, followed by a detailed discussion of the nonlinear interaction
properties of the O-type and P-type 2-soliton solutions. We conclude the paper with a
brief summary and possible significance of the results.

\section{The KP $\tau$-function}
The most convenient representation for the line-soliton solutions is via
the $\tau$-function which plays an important role in the mathematical theory
of the KP equation~\cite{S:81, H:04}. The solution $u(x,y,t)$ of \eqref{kp}
can be expressed as
\begin{equation}
u(x,y,t) = 2(\ln \tau)_{xx}\,, 
\label{tau}
\end{equation}
where $\tau(x,y,t)$ is the KP $\tau$-function which is defined up to an exponential
factor that is linear in $x,y$ and $t$. 

\subsection{The Wronskian form of the $\tau$-function}
The $\tau$-function can be expressed as the Wronskian
determinant~\cite{S:81, FN:83, MS:91}
\begin{equation}
\tau(x,y,t)= \mathrm{Wr}(f_1,\dots,f_N)=
  \begin{pmatrix}
     f_1 & f_2 & \cdots & f_N \\
     f_1^{(1)} & f_2^{(1)} &\cdots & f_N^{(1)} \\
     \vdots & \vdots & & \vdots \\
     f_N^{(N-1)} & f_2^{(N-1)} & \cdots &f_N^{(N-1)}
  \end{pmatrix}\,,
\label{Wr}
\end{equation}
where $f^{(i)}$ denotes the $i^{\mathrm th}$ partial derivative with respect to $x$,
and where the functions $\{f_n\}_{n=1}^N$ form a set of linearly independent solutions 
of the linear system
\begin{equation}\label{fpde}
f_y= f_{xx}\,, \qquad f_t= f_{xxx}\,.
\end{equation}
A remarkable fact is that the KP equation simply turns into a determinant 
identity if one substitutes \eqref{tau} into \eqref{kp} and uses the Wronskian
form of $\tau(x,y,t)$ given by \eqref{Wr} (see e.g., \cite{FN:83, H:04,CK:09} for details). 
This implies that {\it any} linearly
independent set of solutions of the linear system \eqref{fpde} will give rise to a 
solution of the KP equation. Hence it is possible to generate a large class of 
solutions in this way. In particular, the line-soliton solutions are obtained
from the choice
\begin{equation}
f_n(x,y,t)= \sum_{m=1}^{M} a_{nm}\,e^{\theta_m}\,, \quad 
n = 1,2, \ldots, N \,, 
\label{f}
\end{equation}
where $\theta_m(x,y,t) = k_mx +k_m^2y+k_m^3t+\theta_{0m}$ with distinct real 
parameters:\, $k_1 < k_2 < \ldots < k_M$ and real constants 
$\{\theta_{0m}\}_{m=1}^M$. The coefficients $(a_{nm})=:A$ define an 
$N \times M$ constant matrix which is of rank $N$ due to the linear independence 
of the functions $\{f_n\}_{n=1}^N$. Then \eqref{Wr} can be expressed as
\begin{equation}
\tau(x,y,t)= \hspace{-0.3in}\sum_{1\le m_1<\dots<m_N\le M}
  \hspace{-0.3in}A(m_1,\dots,m_N)
  \,\,\,
  \exp[\,\, \theta(m_1,\ldots,m_N) \,]
  \!\!\prod_{1\le s < r\le N}(k_{m_{r}}-k_{m_{s}})\,,
\label{tauexp}
\end{equation}
by expanding the Wronskian determinant using Binet-Cauchy formula. In above,
$\theta(m_1,\ldots,m_N):= \theta_{m_1}+\theta_{m_2}+\ldots+\theta_{m_N}$,
and $A(m_1,\dots,m_N)$ is the maximal minor, i.e., the determinant
of the $N\times N$ sub-matrix of $A$ obtained from columns $1\le m_1<\dots<m_N\le M$.
Note that the $N$ linearly independent rows of the coefficient matrix $A$ span
an $N$-dimensional subspace $W$ of $\Real^M$ so that $A$ can be regarded as a
point of the real Grassmannian Gr$(N,M)$. A different choice of basis for $W$
amounts to performing row operations:\, $A \to GA, \,\, G \in $ GL$(N, \Real)$,
which changes the $\tau$-function in \eqref{tauexp} simply by a scale
factor:\, $\tau \to \det(G)\tau$ but leaves the KP solution $u = 2 (\ln \tau)_{xx}$
invariant. Consequently the coefficient matrix $A$ can be canonically chosen in
the reduced row echelon form (RREF) which has a distinguished set of
pivot columns such that the restriction of $A$ to this set corresponds 
to the $N \times N$ identity matrix. Given any $N \times M$ matrix $A$ of rank $N$, 
its unique RREF gives a standard coordinate for the corresponding point in Gr$(N,M)$.
It is in this sense that the space of all solutions of KP generated by the $\tau$-function
given by \eqref{tauexp} can be identified with the real Grassmann manifold Gr$(N,M)$.
In general, such solutions are not all line-solitons as they can be singular 
where the $\tau$-function becomes zero. The regularity of the solutions in the
entire $xy$-plane and for all values of $t$ is achieved by imposing a further restriction
on the coefficient matrix $A$, namely that all of its $N \times N$ maximal minors 
be non-negative. Such matrices are called {\it totally non-negative} matrices, they
represent the totally non-negative Grassmannian Gr$^+(N,M)$ which forms a closed subset
in Gr$(N,M)$, and which identifies the space of all line-soliton solutions of 
the KP equation.

Some simple examples of $\tau$-functions corresponding to some of the 
solutions mentioned in Section 1 are listed below.

{\it One-soliton solution}:\, In this case, one chooses a single function
in \eqref{f} of the form $f_1 = e^{\theta_1} + e^{\theta_2}$. The coefficient
matrix is simply $A= (1 \quad 1)$, and $\tau = f_1$. Then equation \eqref{tau}
yields the one-soliton solution given by \eqref{onesoliton} in Section 1
with $\theta_0 = \theta_{02}-\theta_{01}$. Note that since $k_1<k_2$,
$\tau \sim e^{\theta_1} \gg  e^{\theta_2}$ and 
$u(x,y,t) \sim 0$ as $x \to -\infty$. 
A similar argument shows that $u$ is also exponentially small as $x \to \infty$.
In this case, one finds from the exact expression in \eqref{onesoliton} that the
solution is localized in the $xy$-plane along the line $\theta_1 = \theta_2$ 
(for fixed $t$) where the two exponentials in the $\tau$-function are precisely 
in balance (see Figure 1(a)). It turns out that even for general line-solitons, 
the solution $u(x,y,t)$ is
localized along certain lines in the $xy$-plane where exactly two exponential terms 
in the $\tau$-function of equation \eqref{tauexp} are in balance,
and they dominate over all other exponential terms. This principle of {\it dominant balance}
was implemented in Refs~\cite{BK:03, K:04, BC:06, CK:08} to identify the asymptotic 
line solitons $[i,j]$ associated with each line-soliton solution of the KP equation.

{\it Miles Y-shape solution}:\, The $\tau$-function for this solution is a sum
of three exponential functions, i.e., 
$ \tau = f_1 = e^{\theta_1}+e^{\theta_2}+e^{\theta_3}$ with the coefficient
matrix $A= (1 \quad 1 \quad 1)$. Applying the dominant balance principle mentioned
above, it is possible here to determine the dominant exponentials and analyze the 
structure of the solution in the $xy$-plane. One finds that the solution $u(x,y,t)$ 
is localized along three lines corresponding to the line soliton $[1,3]$ for $y \gg 1$,
and line solitons $[1,2], \, [2,3]$ for $y \ll 1$ as illustrated in Figure 1(b). 
For $i,\,j \in \{1,2,3\}$, each line soliton 
$[i,j], \, i< j$ is locally of the form of the one-soliton solution \eqref{onesoliton}
with distinct parameters $k_i, \, k_j$. As mentioned in Section 1, the Miles solution
represents a resonant solution of three line solitons. The resonant condition among 
those three line-solitons is given by
\begin{align*}
{\bf k}_{13}={\bf k}_{12}+{\bf k}_{23},\qquad \omega_{13}=\omega_{12}+\omega_{23}\,,
\end{align*}
where ${\bf k}_{ij} = \sf{1}{2}(k_j-k_i,k_j^2-k_i^2)$ and $\omega_{ij} = \sf{1}{2} (k_j^3-k_i^3)$ 
are respectively, the wave vector and frequency of the line soliton $[i,j]$, satisfying
the soliton dispersion relation given in Section 1. 

{\it O-type 2-soliton solution}:\, Here the $\tau$-function
is a Wronskian of two linearly independent solutions of \eqref{fpde}. These solutions
are obtained from \eqref{f} by choosing $N=2, \, M=4$, and the coefficient matrix
$$
A = \begin{pmatrix} 1 & 1 & 0 & 0 \\ 0 & 0 & 1& 1 \end{pmatrix}\,, 
$$
such that $f_1 = e^{\theta_1} + e^{\theta_2}$ and $f_2 = e^{\theta_3} + e^{\theta_4}$.
Note that $A$ is a totally non-negative matrix whose $2 \times 2$ maximal minors
are:\, $A(1,2) = A(3,4) = 0$, and $A(1,3)=A(1,4)=A(2,3)=A(2,4) = 1$.
From \eqref{tauexp}, the resulting $\tau$-function is given by
$$ \tau = (k_3-k_1)e^{\theta(1,3)} + (k_4-k_1)e^{\theta(1,4)} + (k_3-k_2)e^{\theta(2,3)}
+ (k_4-k_2)e^{\theta(2,4)} \,, \qquad \theta(i,j) := \theta_i+\theta_j \,.
$$
Figure 1(c) illustrates the regions in the $xy$-plane where each of the four exponential
terms in the above $\tau$-function are dominant. The line solitons corresponding to
the solution $u = 2(\ln \tau)_{xx}$ are $[1,2]$ and $[3,4]$ for $|y| \to \infty$;
these are localized along the directions where a pair of dominant exponential terms are in
balance.

\subsection{The Grammian form of the $\tau$-function}
The Grammian construction of the line-soliton solutions will be described next.
This form of the solution arises from the so called binary Darboux transformation
of the KP equation (see e.g.,~\cite{MS:91}). In this case, the $\tau$-function 
can be expressed as a a determinant of an $N \times N$ matrix as follows:
\begin{equation}
\tau = \det(\Omega)\,, \qquad 
\Omega_{ij} = \int_{\Gamma}f_ig_j\,dx 
{\color{red}\mathbf{-}} (f_ig_{jx} - f_{ix}g_j)\,dy\,, \quad i,j=1,2,\ldots,N \,,
\label{bdt}
\end{equation}
where $\{f_i\}_{i=1}^N$ solve \eqref{fpde} and $\{g_j\}_{j=1}^N$ solve the
(formal) adjoint system of \eqref{fpde}, namely,  
\begin{equation*}
-g_y = g_{xx}\,, \qquad g_t = g_{xxx} \,.
\label{gpde}
\end{equation*}
It is easily verified from \eqref{fpde} and its adjoint
that $d\Omega_{ij}$ is an exact differential, hence the line 
integral above can be evaluated along any suitable curve $\Gamma$ in the
$xy$-plane such that the integral converges.
Like the Wronskian form, it can be shown that the $\tau$-function in \eqref{bdt}
yields a solution of the KP equation via \eqref{tau} for any choice of the sets
of functions $\{f_i\}_{i=1}^N$ and $\{g_j\}_{j=1}^N$~\cite{H:04}. For the line-soliton
solutions, these are chosen as linear combinations of exponentials (cf. \eqref{f}),
\begin{equation}
f_i = \sum_{n=1}^r b_{in}e^{\phi(p_n)}\,, \qquad \qquad 
g_j = \sum_{n=1}^N \widehat{b}_{jn}e^{-\phi(q_n)}\,, \quad i,j = 1,\ldots,N\,, 
\label{fg}
\end{equation}
where $\phi(k):= kx+k^2y+k^3t+\phi_0(k)$, and all the parameters 
$p_1, \ldots, p_r,\, q_1, \ldots, q_N$ are distinct and real.
Note that $N \neq r$, in general. Thus, $B := (b_{mn})$ is an $N \times r$ matrix 
whereas $\widehat{B} := \widehat{b}_{mn}$ is an invertible, $N \times N$ matrix.
Substituting \eqref{fg} into the matrix elements $\Omega_{ij}$ in \eqref{bdt}, and 
evaluating the line integral along a path $\Gamma$ from 
$({\color{red} \mathbf{-}}\sigma \infty,\, y)$ to $(x,y)$ parallel to the $x$-axis, yields
\begin{equation}
\Omega_{ij}(x,y,t) = c_{ij}+\int_{\Gamma} f_ig_j\, dx = 
c_{ij}+\sum_{m=1}^r\sum_{n=1}^N b_{im}\widehat{b}_{jn}  
\frac{e^{\phi(p_m)-\phi(q_n)}}{p_m-q_n}\,,
\label{Omega}
\end{equation}
where $\sigma=\mathrm{sgn}(p_m-q_n) = \pm 1$ and $c_{ij}$ are arbitrary constants. 
Choosing the constants $c_{ij} = \delta_{ij}$, where $\delta_{ij}$ is the Kr\"onecker 
symbol ($\delta_{ij} = 0,\, i \neq j,\, \delta_{ii} = 1$), ensures
that the matrix $\Omega$ is of rank $N$ so that $\tau \ne 0$. Using \eqref{Omega},
the $\tau$-function in \eqref{bdt} can be expressed in the form
\begin{equation}
\tau = \det(I + BF\widehat{B}^T) = \det(I + CF)\,, 
\label{gram}
\end{equation}
where $C = \widehat{B}^TB$ is an $N \times r$ matrix of constant coefficients 
and $F$ is an $r \times N$ matrix whose entries are given by
$$
F_{mn} = \frac{e^{\phi(p_m)-\phi(q_n)}}{p_m-q_n}\,, \quad m = 1,2,\ldots,r, \quad
n=1,2,\ldots,N\,.  $$
The Grammian form given by the first equality in \eqref{gram} is not unique as
it is possible to obtain the same $\tau$-function from a different choice of 
the matrices $B$ and $\widehat{B}$. In particular, $\widehat{B}$ need not be a square 
matrix although there is always a canonical choice of an $N \times N$ matrix
$\widehat{B}$ that leads to the second equality in \eqref{gram}.
We will not discuss the details of the gauge freedom underlying the
choice of $B$ and $\widehat{B}$ in this article.

Equation \eqref{gram} is the canonical Grammian form for the 
line-soliton solutions, special
cases of which arise from several direct methods of constructing solutions including 
the Hirota method~\cite{H:04}, direct linearization~\cite{AC:91} and dressing 
techniques~\cite{ZS:74}. For example, the O-type $N$-soliton solution is obtained
by setting $r=N$ and $C=I$, the latter being 
the $N\times N$ identity matrix. In this case the
resulting $\tau$-function $\tau = \det(I+F)$ is positive for all $x, y, t$ if
the parameters are ordered as $p_N < p_{N-1} < \ldots < p_1 < q_1 < q_2 < \ldots < q_N$.
Then the corresponding line-soliton solution is non-singular.
More general line-soliton solutions can also be generated from the Grammian
form \eqref{gram} by making appropriate choices for the matrix $C$.
However, in comparison with the Wronskian form, here 
it is less clear how to impose regularity conditions on the obtained solution.
Recall that in the Wronskian construction the ordering $k_1<k_2< \ldots <k_M$
of the parameters and the totally non-negative coefficient matrix $A$ guarantee 
that the KP line-soliton solutions are regular. But no such clear prescription
to obtain regular solutions is known for the Grammian case. 
This issue can be resolved by establishing an equivalence between the Wronskian and 
the Grammian forms for the KP $\tau$-function. This will be done below. 
More specifically, it will be shown by explicit construction that there is a one-to-one
correspondence between the formulas \eqref{tauexp} and \eqref{gram}.

\subsection{Equivalence of the Wronskian and Grammian forms of the $\tau$-function}
It is convenient to first start with the Wronskian form and derive the
$\tau$-function formula \eqref{gram} from that. It follows from either \eqref{tauexp} or 
\eqref{Wr} that the $\tau$-function can be written as:\, $\tau = \det(AEK)$, where
$E=\mathrm{diag}(e^{\theta_1}, e^{\theta_2}, \ldots, e^{\theta_M})$ 
is an $M \times M$ diagonal matrix and $K :=(K_{ij}) = (k_i^{j-1})$ is an $M \times N$ 
Vandermonde matrix. In the following, the $N \times M$ coefficient matrix $A$ will 
be chosen in RREF such that it can be represented as $A= (I\,, J)P$ with $I$ and $J$
denoting respectively, the $N \times N$ and $N \times (M-N)$ sub-matrices of the pivot 
and non-pivot columns of $A$, and $P$ being the $M \times M$ permutation matrix which
shuffles those columns to form $A$. For example, the coefficient matrix for the O-type
2-soliton $\tau$-function discussed above can be represented as
$$
A = \begin{pmatrix} 1 & 1 & 0 & 0 \\ 0 & 0 & 1& 1 \end{pmatrix} =
\begin{pmatrix} 1 & 0 & 1 & 0 \\ 0 & 1 & 0& 1 \end{pmatrix} P_{23} \,,
$$
where $P_{23}$ is the $4 \times 4$ matrix permuting the second and the
third columns. Using the above form of $A$, the determinant for the $\tau$-function becomes
\begin{equation*}
\tau = |AEK| = |(I, J)(PEP^{-1})PK| = \left|(I, J)\begin{pmatrix} E_1 & 0 \\ 0 & E_2 \end{pmatrix}
\begin{pmatrix} K_1 \\ K_2 \end{pmatrix}\right| = |IE_1K_1+JE_2K_2|\,,
\end{equation*}
where $E_1$ and $E_2$ are respectively, $N \times N$ and $(M-N) \times (M-N)$ block
diagonal matrices whose elements are permutations of the set 
$\{e^{\theta_1}, e^{\theta_2}, \ldots, e^{\theta_M}\}$. Similarly, $K_1$ and $K_2$ are
respectively, $N \times N$ and $(M-N) \times N$ matrices obtained by permuting the
rows of the Vandermonde matrix $K$ by $P$. It should be clear from above that in effect,
the $M \times M$ matrix $P$ induces a permutation $\pi$ of the ordered set 
$\{k_1, k_2, \ldots, k_M\}$ which can be expressed as
\begin{equation}
\pi(\{k_1, k_2, \ldots, k_M\}) = \{q_1, q_2, \ldots, q_N, p_1, p_2, \ldots, p_{M-N}\}\,,
\label{permutation}
\end{equation}
after renaming the elements of the permuted set. Accordingly, the matrices $E_1, E_2, K_1, K_2$ 
are redefined as 
\begin{align*}
&E_1 = \mathrm{diag}(e^{\phi(q_1)}, e^{\phi(q_2)}, \ldots, e^{\phi(q_N)})\,, \qquad
E_2 = \mathrm{diag}(e^{\phi(p_1)}, e^{\phi(p_2)}, \ldots, e^{\phi(p_{M-N})})\,, \\
& K_1:= (K_{1ij})_{i,j=1}^N = (q_i^{j-1})\,,  \quad \qquad \qquad 
K_2:= (K_{2ij}) = (p_i^{j-1})\,,  
\end{align*}
for $i=1,\ldots, M-N, \,\, j=1,\ldots, N$, and where $\phi(k)$ is
defined below \eqref{fg}. 
Since the parameters $q_1, \ldots, q_N$ are distinct, the
Vandermonde matrix $K_1$ is invertible. Then the above determinant expression
for $\tau$ can be further manipulated as 
\begin{equation*}
\tau = |E_1K_1|\widehat{\tau}\,, \qquad \qquad \widehat{\tau}=|I+JE_2K_2K_1^{-1}E_1^{-1}|\,,  
\end{equation*}
where $\tau$ and $\widehat{\tau}$ differ by an exponential factor linear in
$x, y, t$, hence both generate the same KP solution via \eqref{tau}. To show that
$\widehat{\tau}$ is indeed the Grammian form of the $\tau$-function, one employs
the following matrix identity which is derived in the Appendix,  
\begin{equation}
K_2K_1^{-1} = D_2\chi D_1^{-1} \,,
\label{vander}
\end{equation}
where the matrices $D_1, D_2$ and the Cauchy matrix $\chi$ are defined as follows:
\begin{align*}
& D_1=\mathrm{diag}\Big(\prod_{j=1, j\neq i}^N(q_i-q_j)\Big)_{i=1}^N\,,
\qquad \qquad D_2 = \mathrm{diag}\Big(\prod_{j=1}^N(p_i-q_j)\Big)_{i=1}^{M-N}\,, \nonumber \\
& \chi = (\chi)_{ij} = \frac{1}{p_i-q_j}\,, \quad i=1,\ldots,M-N, \,\, j=1\ldots,N \,.
\end{align*}
Substitution of \eqref{vander} into the expression for 
$\widehat{\tau}$ above, yields
\begin{equation*}
\widehat{\tau} = |I+JD_2E_2 \chi E_1^{-1}D_1^{-1}|
= |I+(D_1^{-1}JD_2)(E_2 \chi E_1^{-1})| \,.
\end{equation*}
Finally, by setting $r=M-N$ and the $N \times M-N$ matrix 
$C = D_1^{-1}JD_2$, the Grammian form of the $\tau$-function in \eqref{gram} is readily 
recovered from the expression of $\widehat{\tau}$ above.

It is relatively straightforward to reverse the steps described above to obtain the
Wronskian data, i.e., the coefficient matrix $A$ and the ordered set of
parameters $k_1 < k_2 < \ldots < k_M$ from the Grammian data, which consists of the $N \times M-N$ 
matrix $C$ and the {\it unordered} set of parameters $\{q_1, \ldots q_N, p_1, \ldots p_{M-N}\}$.
The key step is to recover the permutation $\pi$ in \eqref{permutation} by simply {\it ordering}
the set of Grammian parameters, i.e.,
$$ \pi^{-1}(\{q_1, \ldots q_N, p_1, \ldots p_{M-N}\}) = \{k_1, k_2, \ldots, k_M\} \,.$$
This in turn, provides the $M \times M$ permutation matrix $P$ from $\pi$, and the matrix
$A$ is then constructed by the formula $A = (I\,, J)P$
with $J = D_1CD_2^{-1}$, thus establishing the 
Grammian-Wronskian equivalence for the $\tau$-function for the KP line-soliton solutions.
Such equivalence was also derived in Ref.~\cite{FN:83} for
special $N$-soliton solutions using similar techniques.

An alternative approach to obtain the Wronskian form from the Grammian in \eqref{bdt}
that is applicable when at least one of the functions $f_i$ or $g_j$ is an exponential, 
was discussed in \cite{GPS:95}. In this case, one can for instance, choose the
functions $\{f_i\}_{i=1}^N$ as $N$ linearly independent solutions of \eqref{fpde},
and $\{g_j:=e^{-\theta_j}\}_{j=1}^N$ with $\theta_j = k_jx+k_j^2y+k_j^3t$ and $k_j > 0$.
Using these forms of $f_i$ and $g_j$ in \eqref{bdt}, and choosing $\Gamma$ to be the
path from $(-\infty, y)$ to $(x,y)$ parallel to the $x$-axis, the integral for the matrix 
element $\Omega_{ij}$ can be expressed as an infinite series in inverse powers of the 
parameters $k_j$ by repeated integration by parts, provided that 
$f_i^{(n)}g_j \to 0, \,\, n\geq 0$ as $x \to -\infty$. Then by taking the limit 
$k_j \to \infty, \, j=1, \ldots, N$, one can recover the Wronskian form \eqref{Wr} 
from the determinant $\det(\Omega)$.

An immediate consequence of the above equivalence result is that it is now possible
to formulate the regularity condition for the KP line-solitons in a precise fashion
from the Grammian form \eqref{gram} of the $\tau$-function. Given the matrices
$C$ and $P$, a necessary and sufficient condition that the KP solution \eqref{tau}
is non-singular is given by the fact that $A = (I,\,J)P$ be 
a totally non-negative matrix. Of course, this is the same condition on the coefficient 
matrix $A$ in the Wronskian form. If in addition, $A$ satisfies the following 
{\it irreducibility} conditions:\\
(i)\, each column of $A$ contains at least one nonzero element, \\
(ii)\, each row of $A$ in RREF contains at least one nonzero element other than 
the pivot (first non-zero entry),\\
then the number of line solitons as $y \to \pm \infty$ in the solution generated by 
the $\tau$-function in \eqref{tauexp} is determined by the size of the matrix $A$.
Namely, one has $N_-=M-N$ and $N_+ = N$, where $N_{\pm}$ denote the number of line
solitons as $y \to \pm \infty$~\cite{BC:06,CK:08, CK:09}. Further analysis of the 
totally non-negative matrix $A$ leads to the precise identification of the line solitons 
and a comprehensive classification scheme for all line-soliton solutions of 
the KP equation. The latter problem is related to the classification of the non-negative 
cells Gr$^+(N,M)$ of the Grassmannian Gr$(N,M)$ in terms of certain types of permutations 
called {\it derangements}~\cite{CK:08}.

It is worth noting that the equivalence between the Grammian and Wronskian
forms of the line-soliton $\tau$-function is not unique to the KP equation
but applies also to other equations whose $\tau$-functions are represented in both forms.
Examples include several $2+1$-dimensional differential-difference equations such as the 
2d-Toda, 2d-Volterra and the differential-difference KP equation as well as 
difference-difference equations such as the fully discrete 2d-Toda and the discrete KP 
(Hirota-Miwa) equation. These will be discussed in future works.

\section{O- and P-type 2-solitons}
In this section the soliton interactions associated with certain
types of $2$-soliton solutions of the KP equation will be investigated. 
These solutions are characterized by a pair of line solitons as $|y| \to \infty$.
Thus, $N_-=N_+=2$, which implies that $N=2$ and $M=4$. The Wronskian form
of the $\tau$-function for these solutions is obtained from \eqref{tauexp}
in terms of the ordered set $k_1 < k_2 < k_3 < k_4$ of distinct 
real parameters and a $2\times 4$ totally non-negative matrix~$A$ as,
\begin{equation}
\tau(x,y,t)= \sum_{1\le r<s\le4}
  (k_s-k_r)\,A(r,s)\, e^{\theta_r+\theta_s}\,,
\label{tau2}
\end{equation}
where $A(r,s)$ denotes the $2\times2$ minors of the matrix $A$.
Taking into account the irreducibility property mentioned in Section 2, there are 
exactly {\it seven} distinct types of totally non-negative $2 \times 4$ matrices,
which lead to seven different types of 2-soliton solutions of the KP 
equation~\cite{CK:08, CK:09}. In contrast, its (1+1)-dimensional version namely,
the KdV equation has only one kind of 2-soliton solution. This is indicative of
a much richer solution space for the (2+1)-dimensional integrable equations than  
the (1+1)-dimensional ones. 

The particular 2-soliton solutions whose nonlinear interaction properties are
studied here, are the O-type 2-soliton which was mentioned in section 2, 
and another one called the P-type (``P" for {\it physical}), which fits better 
with the physical description of oblique interactions of shallow water waves 
for which the KP equation is a good approximation.
In both of these cases, a pair of line solitons interact to form a X-shape in the
$xy$-plane but the details of the interactions are entirely different. The corresponding
coefficient matrices in RREF are given by
\begin{equation*}
A_{\mathrm O}= \begin{pmatrix} 1 & 1 & 0 & 0 \\ 0 &0 &1 &1 \\ \end{pmatrix}, 
\qquad \qquad \qquad
A_{\mathrm P}= \begin{pmatrix} 1 &0 &0 &-1 \\ 0 &1 &1 &0 \\ \end{pmatrix}\,,
\label{Amatrix}
\end{equation*}
which, from \eqref{tau2}, lead to the $\tau$-functions 
\begin{align*}
\tau_{\mathrm O} = (k_3-k_1)e^{\theta_1+\theta_3}+(k_3-k_2)e^{\theta_2+\theta_3}+
(k_4-k_1)e^{\theta_1+\theta_4}+(k_4-k_2)e^{\theta_2+\theta_4} \,, \\
\tau_{\mathrm P} = (k_2-k_1)e^{\theta_1+\theta_2}+(k_3-k_1)e^{\theta_1+\theta_3}+
(k_4-k_2)e^{\theta_2+\theta_4}+(k_4-k_3)e^{\theta_3+\theta_4} \,.
\end{align*}
Applying the principle of dominant balance mentioned in the examples given in 
Section 2, it is possible to identify the line solitons in the two cases. For the
O-type the line solitons are $[1,2]$ and $[3,4]$ as $y \to \pm \infty$, 
whereas for the P-type,
these are $[1,4]$ and $[2,3]$. Recall that locally, a line soliton $[i, j]$ is given
by \eqref{onesoliton} with parameters $k_i$ and $k_j$.

For both O- and P-type solitons, the exact solution can be computed as
$$
u = 2(\ln \tau)_{xx} = 
\frac{\alpha\,\sech^2\phi_1+\beta\,\sech^2\phi_2+\gamma\,\sech^2\phi_1 \,\sech^2\phi_2}
{1+C\, \mathrm{tanh}\phi_1\mathrm{tanh}\phi_2}\,, $$
with $ \alpha = A_1(1-C^2),\,\beta = A_2(1-C^2)$, and $\gamma = (A_1+A_2)C^2+2C\sqrt{A_1A_2}$.
The expressions for the soliton amplitudes $A_1, A_2$, soliton phases $\phi_1, \phi_2$, and the 
constant $C$ are different for the O- and P-type 2-solitons. 
These are given below.

\subsection{O-type interaction}
For the O-type, the amplitudes of the line solitons $[1, 2]$ and $[3, 4]$ are
$$A_1=\alpha_{12} = \half(k_2-k_1)^2\,, \qquad \qquad A_2=\alpha_{34} = \half(k_4-k_3)^2 \,,$$
the soliton phases are expressed in terms of 
$\theta_j = k_jx+k_j^2y+k_j^3t+\theta_{0j}, \,\, j=1,\ldots,4$, as
$$
\phi_1 = \frac{1}{2}(\theta_2-\theta_1)+
\frac{1}{4}\ln \left[ \frac{(k_4-k_2)(k_3-k_2)}{(k_4-k_1)(k_3-k_1)}\right]\,, \qquad \qquad
\phi_2 = \frac{1}{2}(\theta_4-\theta_3)+
\frac{1}{4}\ln \left[ \frac{(k_4-k_2)(k_4-k_1)}{(k_3-k_2)(k_3-k_1)}\right]\,, 
$$
and the constant $C$ is given in terms of the parameters $k_1, \ldots k_4$ by
$$ C=\frac{1-\sqrt{\Delta_{\mathrm O}}}{1+\sqrt{\Delta_\mathrm{O}}} \,, \qquad \qquad 
\Delta_{\mathrm O} = \frac{(k_3-k_2)(k_4-k_1)}{(k_3-k_1)(k_4-k_2)} 
= 1-\frac{(k_2-k_1)(k_4-k_3)}{(k_3-k_1)(k_4-k_2)} < 1 \,,
$$
such that $0 < C < 1$. An O-type soliton is illustrated in Figure 2.
\begin{figure}
\centering
\includegraphics[scale=0.5]{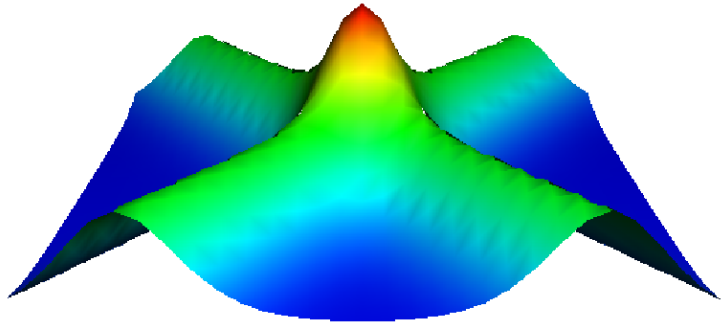}\\
\caption{An O-type 2-soliton. The interaction peak corresponds to an absolute maximum.}
\label{O}
\end{figure}
The function $u(x,y,t)$ has an 
absolute maximum at $\phi_1$ = $\phi_2$ = 0, and the maximum
value at this interaction point is given by (see~\cite{CK:09})
\begin{equation}
u_{\mathrm{max}} = A_1+A_2+2C\sqrt{A_1A_2}\,.
\label{umax}
\end{equation}
\begin{figure}[h!]
\centering
\subfigure[]{
\resizebox*{5cm}{!}{\includegraphics{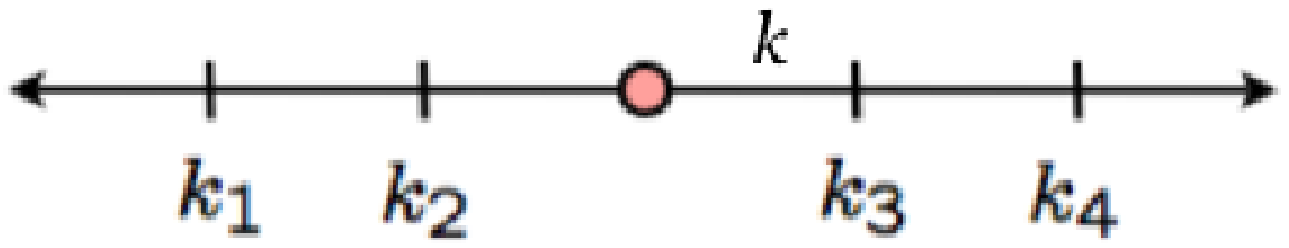}}} \hskip 2.5cm
\subfigure[]{
\resizebox*{5cm}{!}{\includegraphics{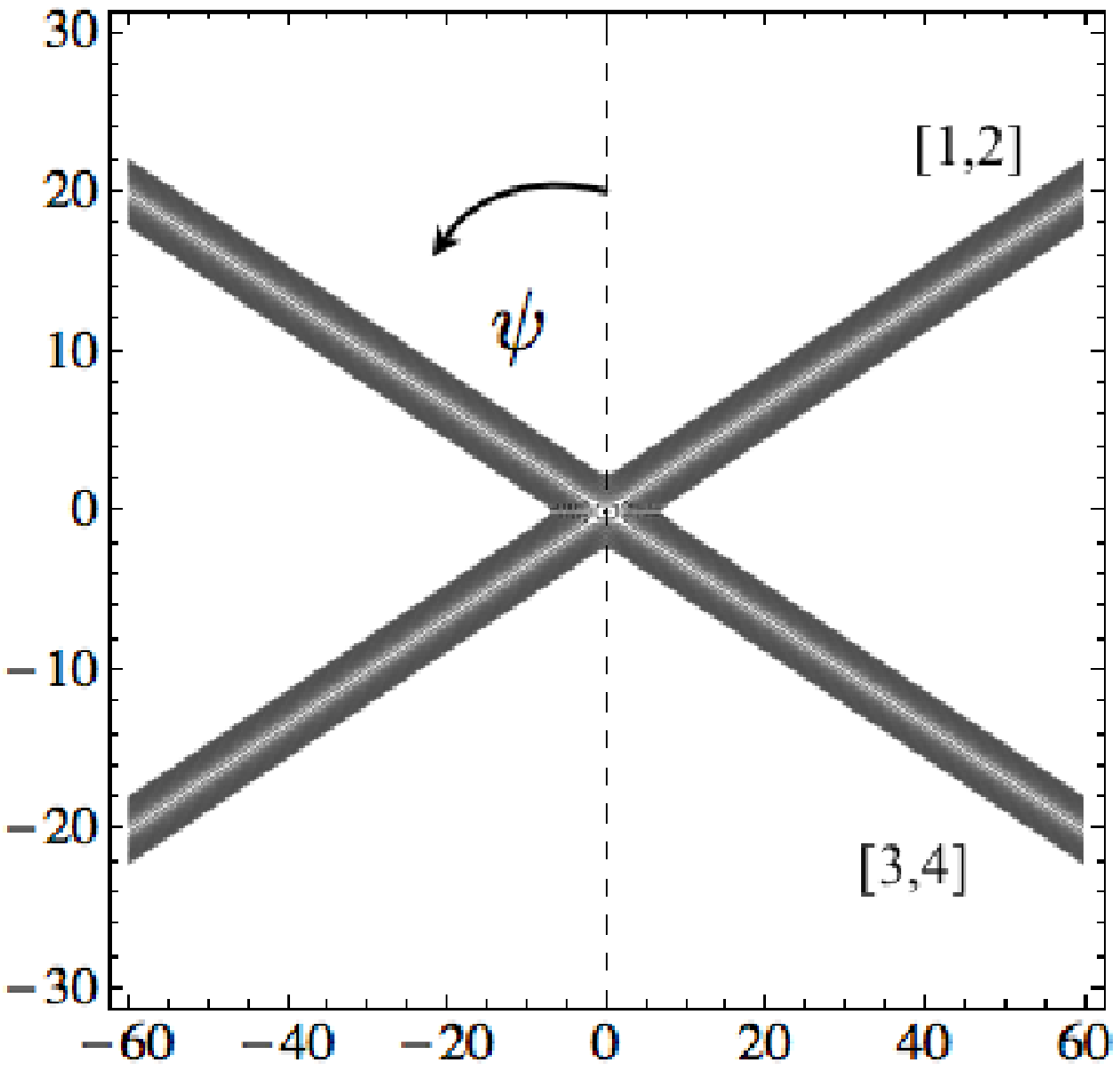}}} 
\caption{(a)~Symmetric choice of the $k$-parameters:\,$k_4=-k_1, k_3=k=-k_2$;
(b)~a O-type 2-soliton with $k_4 = 2=-k_1, \, k_3 = 1 = -k_2$.
Here, each of the $[1,2]$ and $[3,4]$ line soliton makes an angle $\psi$ with the $y$-axis.}
\label{fig3}
\end{figure}
Since $C>0$, the interaction peak $u_{\mathrm{max}}$ is always greater 
than the sum of the individual line soliton amplitudes. In fact, one can 
easily verify that  
$$ A_1+A_2 < u_{\mathrm{max}} < (\sqrt{A_1}+\sqrt{A_2})^2 \,.$$
Furthermore, $u_{\mathrm{max}}$ depends non-linearly on
the incidence angle between the two line solitons. In order to investigate  
this behavior, it is convenient to choose the parameters $k_1, \ldots, k_4$ such
that $k_1 = -k_4$ and $k_2 = -k_3 = -k$ (see Figure 3(a)). 
Then the two line solitons are placed symmetrically about the y-axis, and 
are of equal amplitude, i.e., $A_1 = A_2 = A$.
Denoting the angle between the $[3,4]$-soliton and the $y$-axis by $\psi$ as 
shown in Figure 3(b), it follows that $\tan \psi = (k_3+k_4) = \sqrt{2A}+2k$.
Since $k>0$, the angle $\psi$ is always greater than the critical angle
$$\psi_c = \tan^{-1}(\sqrt{2A}) \,,$$
which depends on the amplitude of the line solitons. Setting
$k_1=-k_4, \, k_2=-k_3$ in the expression for $C$ above, it can be deduced 
from \eqref{umax} that
$$
u_{\mathrm{max}} = \frac{4A}{1+\sqrt{1-s^2}} \,, \qquad\qquad
s = \frac{\tan\psi_c}{\tan\psi} < 1
$$
Therefore, $u_{\mathrm{max}}$ is a decreasing function of the angle $\psi$,
and as $\psi \to \psi_c$ (i.e., $k \to 0$), the peak interaction
amplitude $u_{\mathrm{max}} \to 4A$. Note that in the limit $k=0$, the parameters
$k_2 = k_3$, which implies that the $\tau$-function $\tau_{\mathrm{O}}$ contains only
{\it three} instead of four exponential terms. The resulting solution $u(x,y,t)$ of the
KP equation is a Miles' Y-shape solution corresponding to the confluence of
three line solitons interacting resonantly~\cite{M:77} (see also \cite{CK:09}).
The dependence of $u_{\mathrm{max}}$ with the
angle of interaction is shown in Figure 4 for soliton amplitudes $A=1$ and $A=2$.
Note that in Figure 4(a), $u_{\mathrm max} \to 4$ as $\psi$ approaches
the critical angle $\psi_c \approx 54.7^{\circ}$, while in Figure 4(b), 
$u_{\mathrm max} \to 8$ as $ \psi \to \psi_c \approx 63.4^{\circ}$.
\begin{figure}[h!]
\centering 
\subfigure[]{
\resizebox*{5cm}{!}{\includegraphics{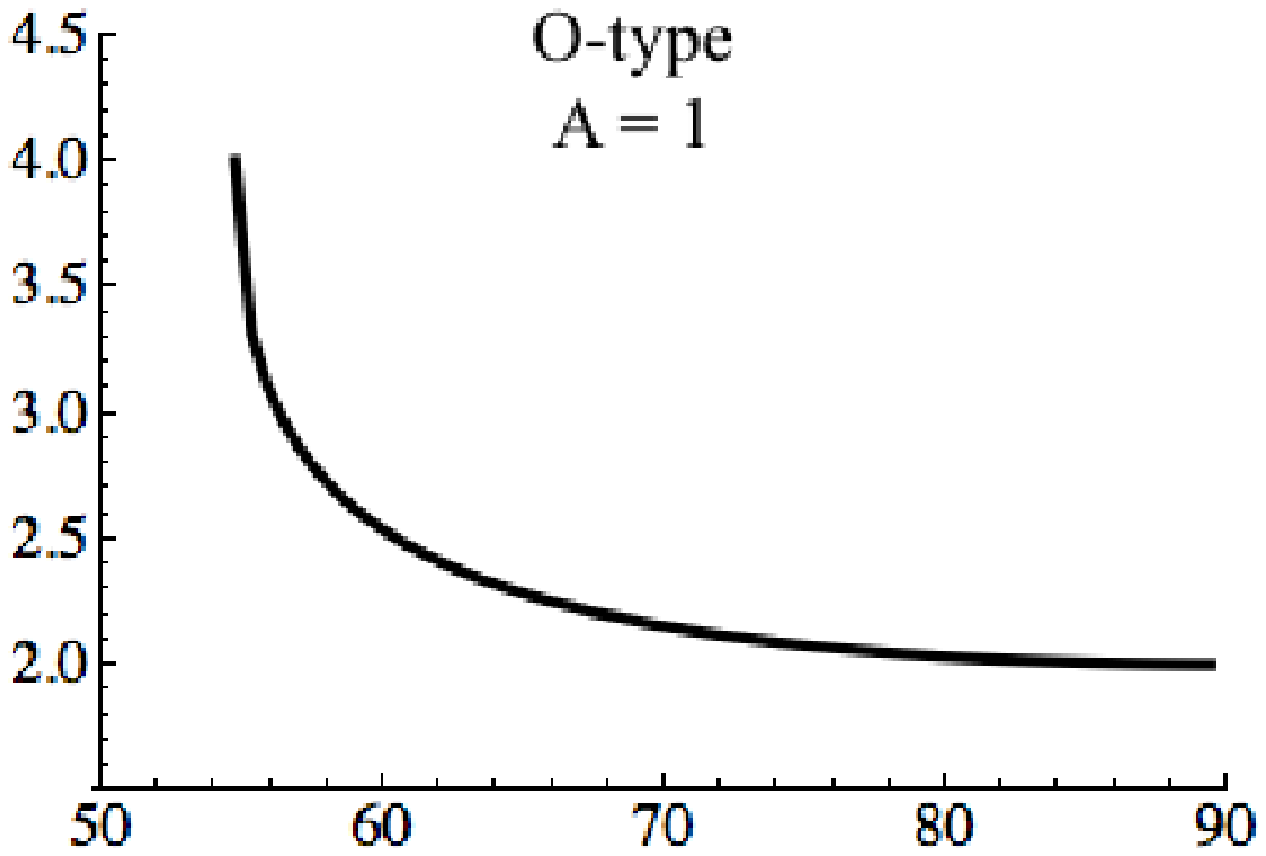}}} \hskip 2.5cm
\subfigure[]{
\resizebox*{5cm}{!}{\includegraphics{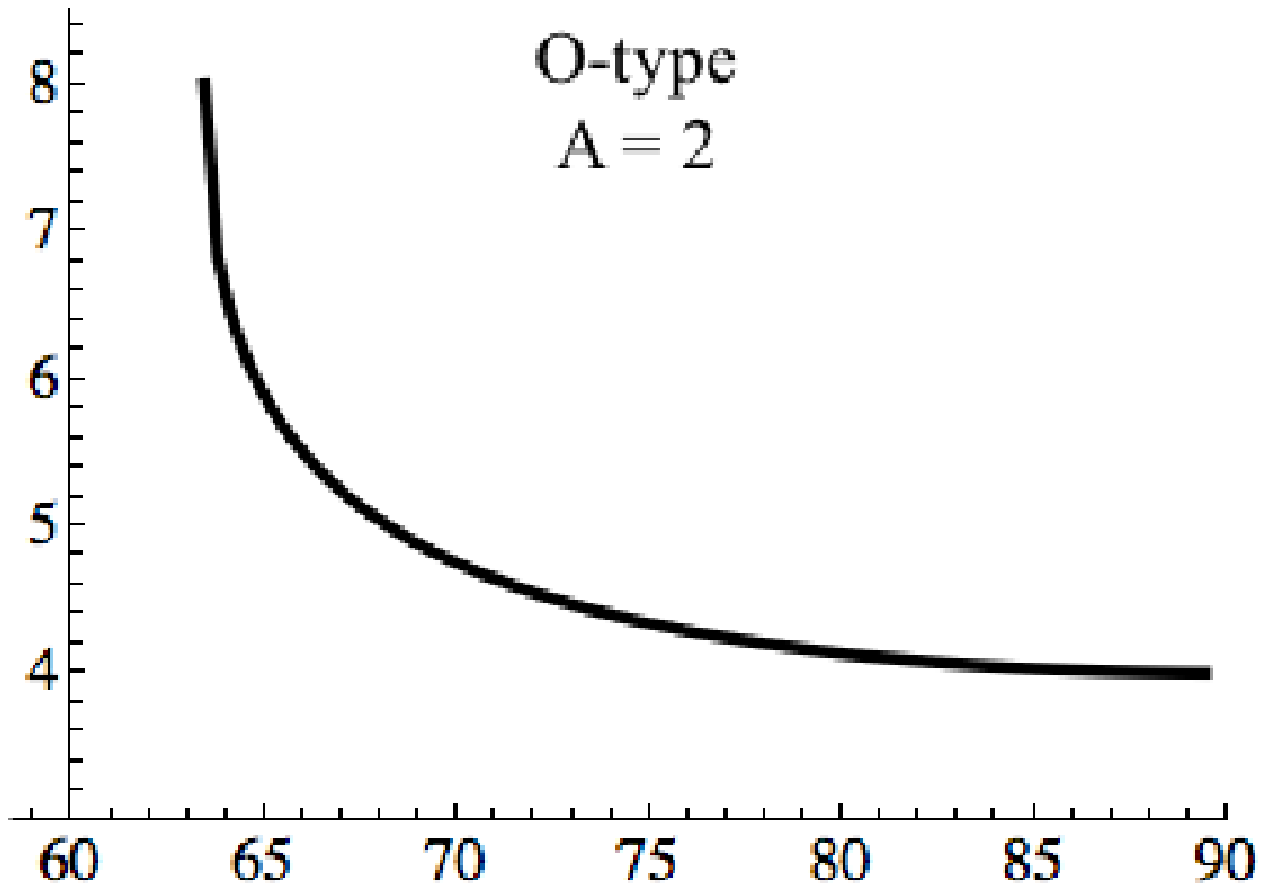}}} 
\caption{Plots of $u_{\mathrm max}$ vs $\psi$ for O-type 2-solitons. The
amplitudes of the line solitons $[1,2]$ and $[3,4]$ are fixed and equal, i.e., 
$A_1=A_2=A$:\, (a)~$A=1,\,\psi_c \approx 54.7^{\circ}$; 
(b)~$A=2,\,\psi_c \approx 63.4^{\circ}$.}
\label{fig4}
\end{figure}

\subsection{P-Type interaction} 
For the P-type, the soliton amplitudes 
$$ A_1=\alpha_{14} = \half(k_4-k_1)^2\,, \qquad \qquad
A_2=\alpha_{23} = \half(k_3-k_2)^2 \,,$$
are always unequal, and $A_1 > A_2$. The soliton phases are given by
$$
\phi_1 = \frac{1}{2}(\theta_4-\theta_1)+
\frac{1}{4}\ln \left[\frac{(k_2-k_1)(k_3-k_1)}{(k_4-k_2)(k_4-k_3)}\right]\,, \qquad \qquad
\phi_2 = \frac{1}{2}(\theta_3-\theta_2)+
\frac{1}{4}\ln \left[ \frac{(k_4-k_2)(k_2-k_1)}{(k_4-k_3)(k_3-k_1)}\right]\,, 
$$
and the constant $C$ is negative since
$$
C = - \frac{1-\sqrt{\Delta_P}}{1+\sqrt{\Delta_P}} \,, \qquad \qquad
\Delta_P = \frac{(k_2-k_1)(k_4-k_3)}{(k_3-k_1)(k_4-k_2)} < 1 \,.
$$
A P-type 2-soliton solution is illustrated in Figure 5.
\begin{figure}
\centering
\subfigure[]{
\resizebox*{5cm}{!}{\includegraphics{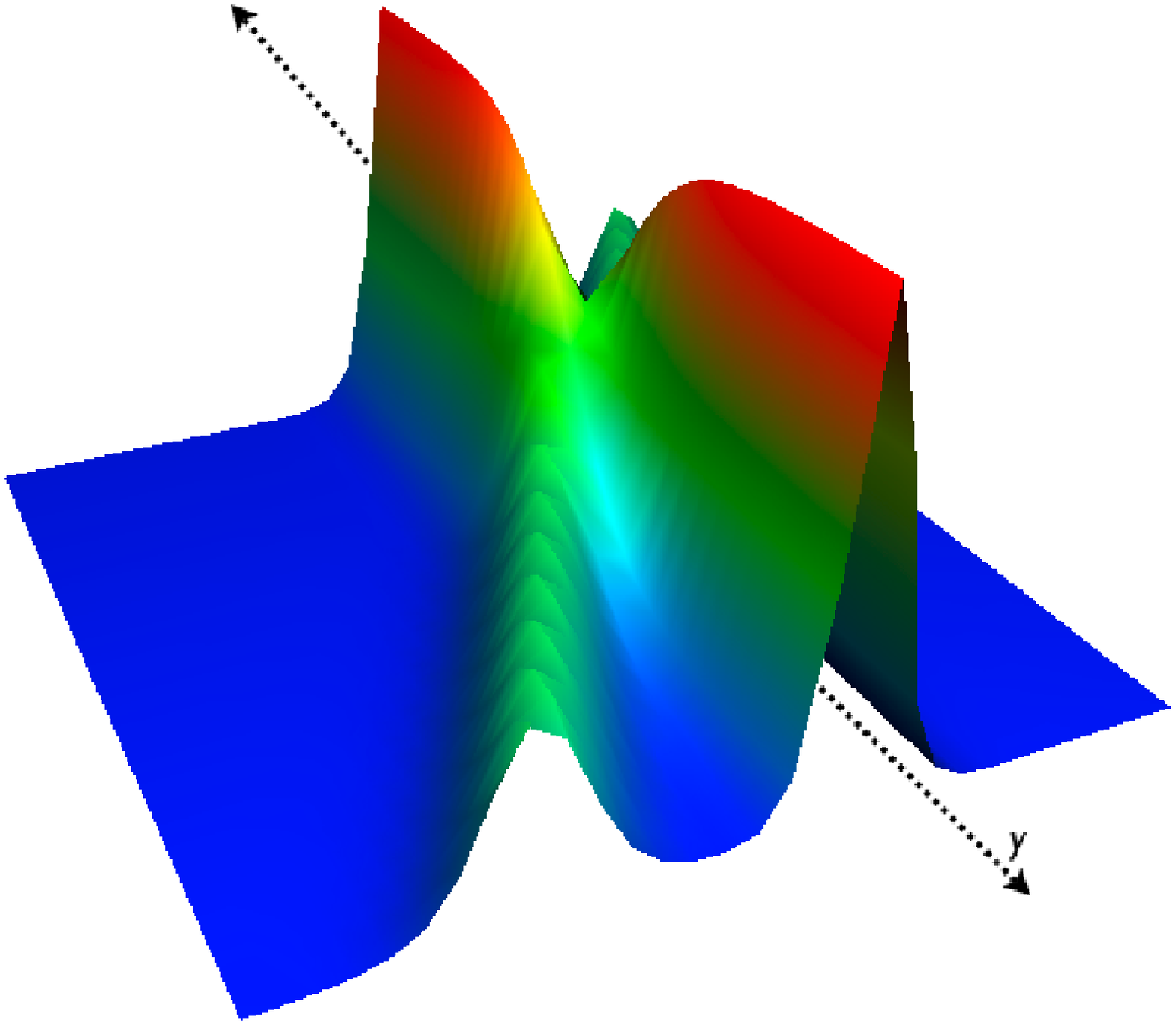}}} \hskip 2.5cm
\subfigure[]{
\resizebox*{5cm}{!}{\includegraphics{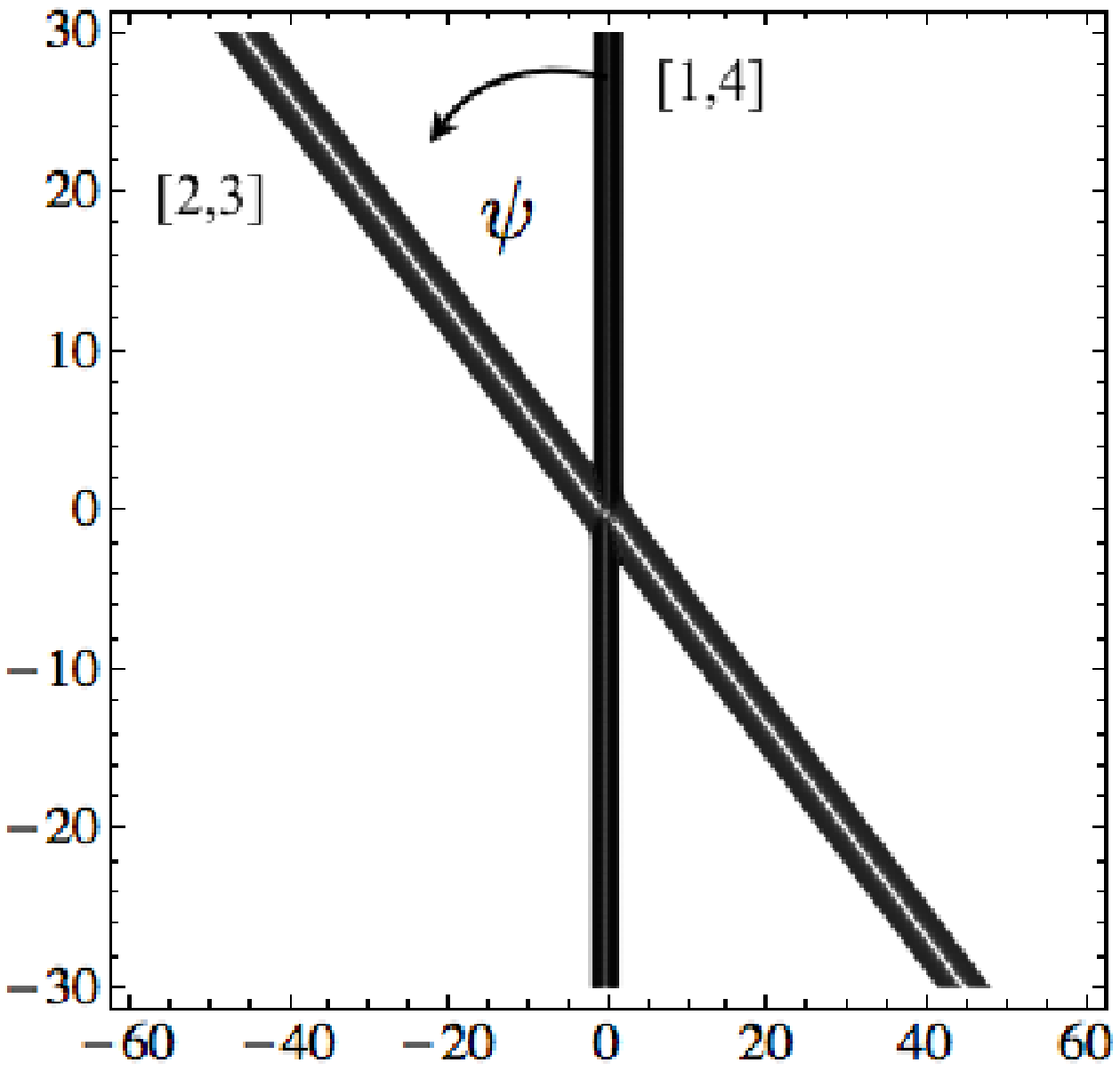}}} 
\caption{(a)\, A P-type 2-soliton. The interaction between the line solitons
is characterized by a saddle point; (b)\, Contour plot of a P-type 2-soliton
with $k_1=-3, k_2= -0.5, k_3 = 2, k_4=3$.
The line soliton $[1,4]$ is along the $y$-axis, the angle between the line solitons
$[1,4]$ and $[2,3]$ is $\psi$.}
\label{P}
\end{figure}
In contrast to the O-type, the P-type 2-soliton solution
$u(x,y,t)$ corresponds to {\it saddle} at the interaction point 
$\phi_1=\phi_2 = 0$. The value of the solution at the interaction point 
is 
\begin{equation}
u_{\mathrm{int}} = A_1+A_2+2C\sqrt{A_1A_2}\,.
\label{uint}
\end{equation}
Since $ C < 0$ for the P-type solution, $u_{\mathrm{int}}$ is always less than 
the sum of the amplitudes of the line solitons $[1,4]$ and $[2,3]$. 
A more precise bound for $u_{\mathrm{int}}$ is obtained as follows~\cite{CK:09}:
From the above expression for $C$, one can calculate
$$
-\,C = \frac{1-\sqrt{\Delta_P}}{1+\sqrt{\Delta_P}} =
\frac{\sqrt{(1-x)(1-y)} - \sqrt{xy}}{\sqrt{(1-x)(1-y)}+\sqrt{xy}}=
\frac{1-(x+y)}{(\sqrt{(1-x)(1-y)}+\sqrt{xy})^2}\,,
$$
where $x=\sf{k_2-k_1}{k_4-k_1}$ and $y=\sf{k_4-k_3}{k_4-k_1}$. 
Since $0 < x, y < 1$, the following inequalities 
$$
\sqrt{xy} \leq \half(x+y)\,, \qquad \qquad \sqrt{(1-x)(1-y)} \leq 1-\half(x+y)
$$
hold, and from these it can be easily deduced that $\sqrt{(1-x)(1-y)}+\sqrt{xy} \leq 1$.
The latter inequality leads to 
$$
\frac{1-\sqrt{\Delta_P}}{1+\sqrt{\Delta_P}} \ge 1-(x+y)=
\frac{k_3-k_2}{k_4-k_1}=\sqrt{\frac{A_2}{A_1}} \,.
$$
Thus, $-1 \leq C \leq -\sqrt{A_2/A_1}$, and from \eqref{uint} it follows that 
$$(\sqrt{A_1}-\sqrt{A_2})^2 < u_{\mathrm{int}} \leq A_1-A_2 \,.$$

Next we investigate the dependence of the amplitude $u_{\mathrm{int}}$
at the interaction point on the angle between the two line solitons $[1,4]$
and $[2,3]$ with fixed amplitudes $A_1$ and $A_2$. For simplicity, we take the 
$[1,4]$-soliton along the $y$-axis by setting $k_1+k_4=0$, and denote by $\psi$ the 
angle between the solitons as in Figure 5(b). Then 
$$\tan \psi = k_2+k_3 = 2(k_2-k_1)-(\sqrt{2A_1}-\sqrt{2A_2}) \,, $$ 
after using $k_1+k_4=0$. Similarly, one calculates
$(k_2-k_1) + (k_4-k_3) = \sqrt{2A_1}-\sqrt{2A_2}$, which implies that 
$0 < k_2-k_1 < \sqrt{2A_1}-\sqrt{2A_2}$. Hence for fixed amplitudes $A_1$ and $A_2$,
the angle $\psi$ between the two line solitons satisfies 
$$ -\psi_c < \psi < \psi_c \,, \qquad \qquad 
\psi_c = \tan^{-1}(\sqrt{2A_1}-\sqrt{2A_2}) \,.$$  
The limiting cases $\psi = \pm \psi_c$ correspond to $k_1=k_2$ or $k_3=k_4$.
Either case leads to a degeneration of the P-type 2-soliton to 
a Y-shape solution similar to O-type 2-soliton situation. 
In terms of $A_1, \, A_2$ and the angle $\psi$, the quantity $\Delta_P$ can be
expressed as
$$ \Delta_P = \frac{(k_2-k_1)(k_4-k_3)}{(k_3-k_1)(k_4-k_2)} = 
\frac {\tan^2 \psi_c - \tan^2 \psi}{(\sqrt{2A_1}+\sqrt{2A_2})^2-\tan^2 \psi} \,, $$
and then from \eqref{uint} one obtains
$$ u_{\mathrm int} = \half \Big[\tan^2 \psi + 
\sqrt{(\tan^2 \psi_c-\tan^2 \psi)((\sqrt{2A_1}+\sqrt{2A_2})^2-\tan^2 \psi)} \Big] \,.   
$$
It follows from the above expression that $u_{\mathrm{int}}$ is an even 
function of the angle $\psi$ between the line solitons $[1,4]$ and $[2,3]$
for $|\psi| < \psi_c$. When $\psi=0$, i.e., when the two line solitons are both
parallel to the $y$-axis, $u_{\mathrm{int}}$ reaches its maximum value of $A_1-A_2$
which is the difference between the amplitudes of the two line solitons.
As $\psi \to \pm\psi_c$, the interaction amplitude approaches its lower
bound, i.e.,  $u_{\mathrm{int}} \to \half\tan^2 \psi_c = (\sqrt{A_1}-\sqrt{A_2})^2$.
Figure 6 below illustrates the plots of
$u_{\mathrm{int}}$ as a function of the angle $\psi$ for fixed values
of soliton amplitudes $A_1$ and $A_2$.
\begin{figure}[h!]
\centering
\subfigure[]{
\resizebox*{5cm}{!}{\includegraphics{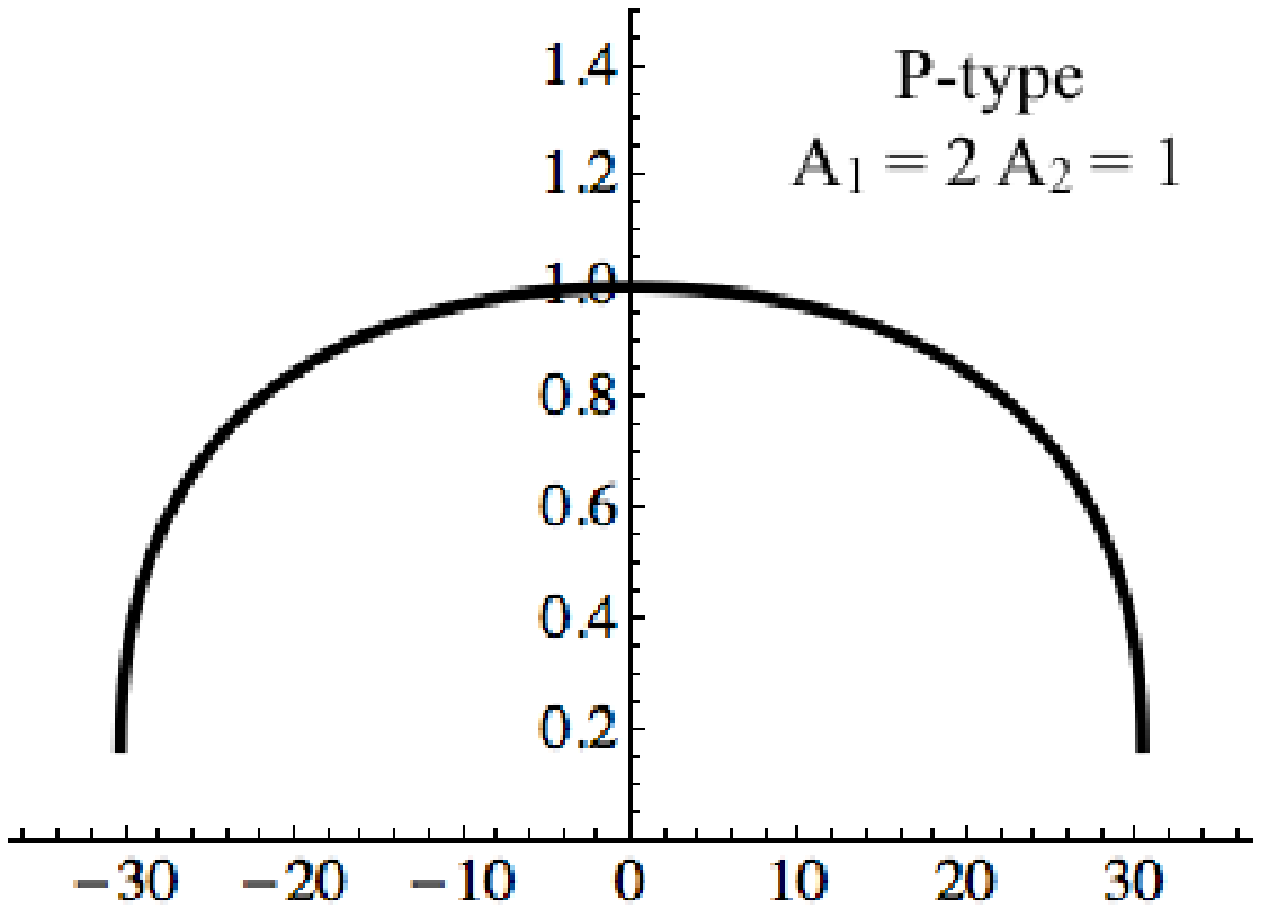}}} \hskip 2.5cm
\subfigure[]{
\resizebox*{5cm}{!}{\includegraphics{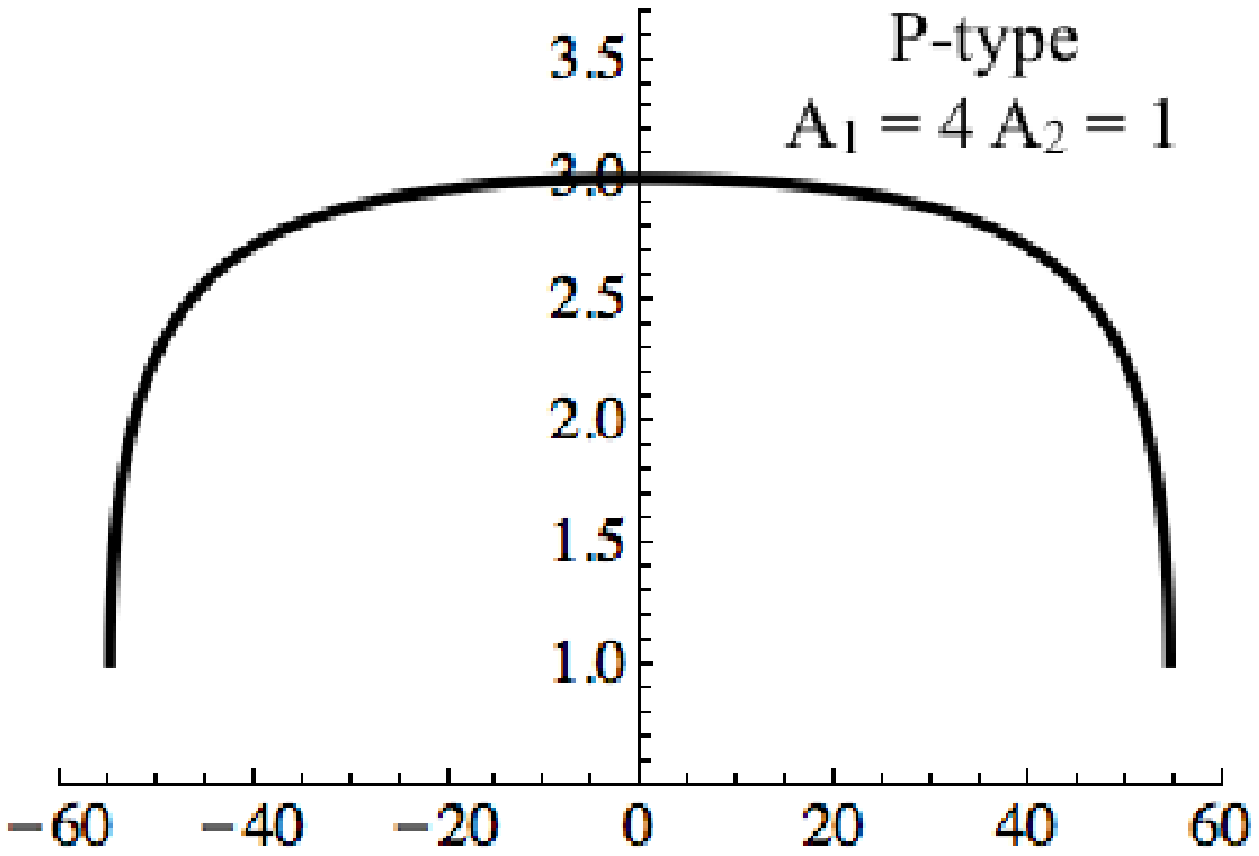}}} 
\caption{Plots of $u_{\mathrm int}$ vs interaction angle $\psi$ for P-type
2-solitons with fixed amplitudes $A_1, \, A_2$ of the line solitons $[1,4]$ and
$[2,3]$:\, (a)~$A_1=2, \, A_1=1$ and $\psi_c \approx 30.4^{\circ}$. The maximum value
of $u_{\mathrm int}$ is 1 at $\psi=0$, and
$u_{\mathrm int} \to (\sqrt{A_1}-\sqrt{A_2})^2 \approx 0.17$ as
$\psi \to \pm \psi_c$; (b)~$A_1=4, \, A_1=1$ and $\psi_c \approx 54.7^{\circ}$.
The maximum interaction amplitude $(u_{\mathrm int})_{\mathrm max} = 3$ at $\psi=0$, and
$u_{\mathrm int} \to 1$ as $\psi \to \pm \psi_c$.}
\label{fig6}
\end{figure}
\section{Conclusion}
In this article we have considered a special class of non-singular solutions of the
KP equation referred to as the line-solitons which decay exponentially 
as $x, y \to \pm \infty$ except along certain directions in the $xy$-plane.
Such solutions exhibit a variety of time-dependent spatial patterns
due to resonant soliton interactions (Y-shape) as well as non-resonant
interactions (X-shape). The exact analytic form of these solutions can be derived 
from the associated $\tau$-functions which are expressible either as Wronskians  
or as Gram determinants. It is remarkable that the KP equation possesses
such a rich structure of line-soliton solutions generated from a simple form of the 
$\tau$-function. It turns out that the solution manifold of the line-solitons
is parametrized by a discrete set of real distinct parameters and the space of totally 
non-negative matrices. This characterization is clear from the Wronskian form of
the KP $\tau$-function but not so transparent from its Grammian form. It is
perhaps due to this difficulty in imposing appropriate regularity conditions
that only a handful of line-soliton solutions were explicitly known via the 
direct algebraic methods which used the Grammian form of the $\tau$-function.
This issue has been resolved in this article where a one-to-one correspondence
between the two forms of the line-soliton $\tau$-function has been established.
Consequently, it is now possible to derive non-singular line-soliton solutions 
using the Grammian form of the $\tau$-function as well.

Another problem discussed in this article is the nonlinear soliton interactions  
for certain types of 2-soliton solutions. In particular, the amplitude of the
nonlinear interaction has been explicitly calculated from the exact analytic 
expression for each of the 2-soliton solutions considered in this paper. Moreover,
the dependence of the interaction amplitude on the angle between the line solitons
has been investigated by keeping the line soliton amplitudes fixed.
One possible physical significance of such results lies in the study of oblique  
nonlinear interactions of weakly 2-dimensional solitary waves in shallow water.
The physical mechanism generating large amplitude waves of extreme
elevations from the interaction of two (or more) smaller 
amplitude solitary waves in shallow water constitutes an important 
open problem. It is believed
that in appropriate parameter regimes the line-solitons of the KP equation can 
serve as a reasonably good test-bed for the description and analysis of nonlinear
solitary wave dynamics. As a qualitative evidence, one may consider the simple example
of the O-type 2-soliton interaction where the interaction peak amplitude may reach
as high as {\it four} times the amplitude of the individual line solitons.

For both O- and P-type solutions, the range of this interaction angle is found to 
be limited by a critical value $\psi_c$ which depends on the amplitudes of the line 
solitons. It is however important to note that the theory of KP line-solitons can 
still be applied to study wave interaction where the angle between the incident 
waves is {\it outside} of the prescribed range for the O- or the P-type soliton 
solutions. In such cases, the wave dynamics is governed by other types of KP 2-soliton 
solutions which have recently been uncovered~\cite{BK:03, BC:06, CK:08}.
The study of the interaction properties of these newly found solutions is a topic 
of future investigation.
\section*{Acknowledgments}
We thank Yuji Kodama for useful discussions. The research of the first and second 
authors is partially supported by the NSF grant DMS-0807404.


\appendix 
\section{}
Here we provide an elementary derivation of the matrix identity \eqref{vander}
involving the Vandermonde matrices $K_1$ and $K_2$. Recall that $K_1$ is an 
$N \times N$ matrix with elements $(K_1)_{ij} = q_i^{j-1}$, and $K_2$ is an 
$N \times (M-N)$ matrix with elements $(K_2)_{ij} = p_i^{j-1}$ where 
$\{q_1,q_2, \ldots, q_N, p_1, p_2, \ldots, p_{M-N}\}$ is a set of distinct elements.

Consider $N$ monic polynomials $P_1(x), P_2(x), \ldots, P_N(x)$ such that each polynomial 
is of degree $N-1$ and has {\it distinct} real roots which consist of $N-1$ elements
from the set $\{q_1, q_2, \ldots, q_N\}$. These polynomials can be expressed as
\begin{equation}
P_j(x) := \prod_{k=1, k \neq j}^N(x-q_k)= \sum_{n=1}^N P_{nj}x^{n-1} \,,
\quad j=1,2, \ldots N\,,
\label{poly} 
\end{equation}
where $P_{nj}$ is the coefficient of $x^{n-1}$ in the polynomial $P_j(x)$, and $P_{Nj}=1$.
An immediate consequence of \eqref{poly} is that the polynomials satisfy
\begin{equation}
P_j(q_i) = \sum_{n=1}^N P_{nj}q_i^{n-1} = (K_1P)_{ij} = (D_1)_{ii}\delta_{ij}\,, \quad
i, k = 1,2, \ldots, N\,,
\label{Pq}  
\end{equation}
where $P = \big(P_{nj}\big)_{n,j=1}^N$ is the matrix of polynomial coefficients,
$D_1$ is the diagonal matrix defined below \eqref{vander}, and $\delta_{ij}$ is 
the Kr\"onecker symbol. Equation \eqref{Pq} gives the matrix equation
$K_1P = D_1$ which implies that 
\begin{equation}
P = K_1^{-1}D_1
\label{PK1} 
\end{equation}
since $K_1$ is invertible. Next, evaluating the polynomials in
\eqref{poly} at $x=p_i$, yields
\begin{equation*}
P_j(p_i) = \sum_{n=1}^N P_{nj}p_i^{n-1} = \prod_{k=1, k \neq j}^N(p_i-q_k)
= \frac{\displaystyle\prod_{n=1}^N(p_i-q_n)}{p_i-q_j}\,, 
\end{equation*}
for $i=1,\ldots, M-N, \,\, j=1, \ldots, N$. These can be represented by the matrix equation 
\begin{equation}
K_2P = D_2\chi \,,
\label{PK2} 
\end{equation}
where the diagonal matrix $D_2$ and the Cauchy matrix $\chi$ are also defined 
below \eqref{vander}. Combining \eqref{PK1} and \eqref{PK2} gives the
desired identity \eqref{vander}.

\end{document}